Title Page

*Classification:* BIOLOGICAL SCIENCES

# Coordinates in low-dimensional cell shape-space discriminate migration dynamics from single static cell images

Xiuxiu He[1], Kuangcai Chen[2], Ning Fang[2], Yi Jiang[1*]

[1] Department of Mathematics and Statistics, Georgia State University, 14th Floor, 25 Park Place, GA, Atlanta, USA 30303-3083.

[2] Department of Chemistry, Georgia State University, 3rd Floor, 145 Piedmont Ave SE, Atlanta, GA, USA 30303.

[*] Corresponding author. Email: yjiang12@gsu.edu.

Text

**Abstract:**

Cell shape has long been used to discern cell phenotypes and states, but the underlying premise has not been quantitatively tested. Here, we show that a single cell image can be used to discriminate its migration behavior by analyzing a large number of cell migration data in vitro. We analyzed a large number of two-dimensional cell migration images over time and found that the cell shape variation space has only six dimensions, and migration behavior can be determined by the coordinates of a single cell image in this 6-dimensional shape-space. We further show that this is possible because persistent cell migration is characterized by spatial-temporally coordinated protrusion and contraction, and a distribution signature in the shape-space. Our findings provide a quantitative underpinning for using cell morphology to differentiate cell dynamical behavior.

**Significance Statement**

Cell biology and pathology have traditionally relied on cell shape to discern cell phenotypes and states. Can we deduce the dynamical behavior of a cell from its shape? We found, based on analyzing $10^6$ cell images of 2D migration, that a static cell shape can discriminate migration patterns: not through simple shape features, but through its coordinates in the low-dimensional shape space. Further analysis explains that the dynamical signatures of intracellular mechanochemical coupling for migration is reflected in the spatiotemporal coordination of cell shape.

**Introduction**

Cell shape has traditionally been an important feature that biologists and pathologists rely on to discern cell type, behavior, and state. The underlying assumption of this practice, however,

Text

has not been systematically and quantitatively tested. Many factors, e.g., the cytoskeleton, the membrane, and adhesion, interact to give rise to cell shape. The detailed molecular scale understanding of these factors is available, but a cellular scale understanding of shape formation is still lacking. In this study, we used cell migration as an example of a dynamical process to address the fundamental question: can a single, static image of cell shape discriminate the dynamical behavior? The shape of a cell changes during migration: the front adheres to the substrate, protrudes outward, and the rear contracts; the number and size of protrusions change as a function of cell state and environment. Most of the molecular details for each individual step and the interaction between protrusion and adhesion have been established (1-3). For motile keratocytes, shape can be used to predict migration speed (4). Cell migration essentially is the sum of all the coordinated changes in cell shape. Genetically identical cells in the same environment exhibit rapid, slow, or negligible locomotion. We analyzed this intrinsic phenotypic migration heterogeneity in mouse fibroblast (NIH3T3) and human glioblastoma (LN229) cells, both are highly motile cell lines. Mouse fibroblast cells have been extensively used as a model system for cell migration (5, 6), making it a good choice for further shape analysis and an ideal comparison with human glioblastoma cells, whose migration is the hallmark of aggressive brain tumor and whose shape and migration behavior has not been quantitatively studied.

**Results**

**Heterogeneous cellular persistence and migration behavior**

Using confocal and epifluorescence live cell imaging, we observed the spontaneous movement of low-density, single NIH3T3 ($n = 517$) and LN229 ($n = 510$) cells plated on fibronectin-coated polyacrylamide gels in the absence of any symmetry-breaking gradients. Cell movements were recorded at a rate of 1 frame per minute for up to 10 hours, resulting in 373,796 single cell images in total. The outline of each cell was segmented (**Fig. 1***A* and **Supplementary**

Text

**Information**). Cell shape (**Fig. 1B**) was represented using a circular map $\Omega(\theta, t)$, which is the distance from the centroid to the cell periphery in the direction θ and at the time t. θ=0 was fixed in the positive direction of the horizontal axis. The dynamics captured include both cell shape deformation (**Fig. 1B**) and the centroid movement. The efficiency of cell migration depends on two essential parameters: cell centroid speed variability and directional persistence (7).

We first need to define migrator and non-migrator. During migration, cells show large variability in simple shape measurements including cell area, aspect ratio, major and minor axes, across the populations of both LN229 and NIH3T3 cells (**Fig. S1**): none of them can distinguish migrator vs non-migrators. To separate cells into migrators and non-migrators, we use the directionality ratio (DR), which is the ratio between the straight-line distance ($d_{n\Delta t}$) and path length ($D_{n\Delta t}$) of the trajectory between the starting position and the current position.

$$DR(n\Delta t) = \frac{d_{n\Delta t}}{D_{n\Delta t}} = \frac{[(x_{n\Delta t}-x_{0\Delta t})^2+(y_{n\Delta t}-y_{0\Delta t})^2]^{\frac{1}{2}}}{\sum_{i=1}^{n}[(x_{i\Delta t}-x_{(i-1)\Delta t})^2+(y_{i\Delta t}-y_{(i-1)\Delta t})^2]^{\frac{1}{2}}}, \quad (1)$$

where $(x_{i\Delta t}, y_{i\Delta t})$ is the cell centroid of the $i^{th}$ frame $(\theta, i\Delta t)$ (**Fig. 1D**); $\Delta t$ is 1 minute in all our data and analysis. DR is an intuitive and effective metric for quantifying directional bias of cell trajectories (**Fig. 1E**). The choice of threshold of DR is optimized based on the analysis of all our cell data (**Fig. S2B**). We found that the mean square displacement (MSD) of the cell trajectories follows a power law $MSD(t) \sim Dt^{\tilde{\alpha}}$, where D is the diffusion coefficient and the exponent α characterizes the nature of diffusive motion. However, for both NIH3T3 and LN229 cells, neither D nor $\tilde{\alpha}$ value can serve as a quantitative concise descriptor of cell motility pattern (**Fig. 1E** and **Fig. S2**), suggesting that diffusivity cannot easily discriminate migrators and non-migrators. It has been shown previously to describe the migration characteristics of fibrosarcoma HT1080 cells (8). For both LN229 and NIH3T3 cells, we found that migrators have greater

Text

persistent time P and smaller cell speed variation S, compared with non-migrators (**Fig. 1F**), which can be explained using the stochastic differential equation that non-migrators have greater magnitude of random fluctuations in speed or less coordination in membrane displacements (**Fig. 1F-G** and **Supplementary Information**).

**Low-dimensional variability of cell shape and cell migration behavior**

To determine how a cell changes its shape during migration, we analyzed the 2D cell shapes $\Omega(\theta, t)$ and the membrane displacement $\mu(\theta, t) = \frac{d\Omega(\theta,t)}{dt}$. The latter captures membrane deformation that varies in size, quantity, and location. The membrane displacements are fewer but larger in size and are more biased in direction for migrators (**Fig. 2A**) than for non-migrators (**Fig. 2B**). The quantity, width and magnitude of directional membrane displacement reflects the intrinsic variability in the underlying molecular mechanisms that result in cell shape differences.

To characterize the cell shape variability quantitatively, we applied principal components analysis (4, 9) to the shapes of NIH3T3 and LN229 cells separately. We found that six orthogonal modes of shape variation (**Fig. 2C** and **Fig. 2D**) account for ~99% of the total shape variability in both cell lines. These shape modes provide a low-dimensional (6-dimensional) space, where each mode is an axis of the coordinate system. The first mode describes the change of cell area through isotropic expansion and contraction - it accounts for the majority of the shape variations for both cell lines. The second mode shows four regions of variation that alternate between contraction and protrusion; the contraction and protrusions are diametrically opposed along the cell periphery, indicating directional cell migration. Modes 3-6 describe shape variations with increasingly larger number of protrusion/contraction regions that are smaller in size. The cell shape $\Omega(\theta, t)$ can be approximated using linear combination of these major modes,

Text

$$\Omega(\theta, t) \approx \sum_{i=1}^{6} \alpha_i(t)\vec{\phi_i}(t) \quad (2)$$

where $\alpha_i(t)$ is the amplitude along the axis $\vec{\phi_i}$. We can then write a stochastic equation of motion for each cell shape that based on the time-lapse images of the whole migration trajectory, in which the parameters connect the cytoskeleton dynamics to the shape determination (**Supporting Information**).

To determine if this shape-space can help to distinguish migrators and non-migrators, we projected every single cell shape in this space, in particular the 2D plane formed by the first and the second modes. The joint probability density, $\rho(\alpha_1, \alpha_2)$, shows the distribution of the coordinates in the plane. For both cell lines, all non-migrators' coordinates $(\alpha_1, \alpha_2)$ are confined in a small area close to origin (**Fig. 2E** and **Fig. 2G**), because non-migrators have low amplitudes in shape variation modes 1 and 2. In contrast, the coordinates of migrators $(\alpha_1, \alpha_2)$ spread much wider, corresponding to higher amplitudes in shape variation modes 1 and 2. The migrators occupy space completely separated from non-migrators in NIH3T3 cells (**Fig. 2F**). In LN229, coordinates of migrators and non-migrators overlaps; however, for migrators, the majority of its frames during their migration process reside outside the non-migration region (**Fig. 2H** and **Fig. S3**). The latter is consistent with the observation that some LN229 cells can switch between migratory and non-migratory states. We compared all the 2D planes in the 6-dimensional shape-space and found that the plane of the modes 1 and 2 offers the best segregation of migrators and non-migrators (Combinations of other shape modes are in **Fig. S4**.). These results suggest that the complex shape variations for mammalian cells can be reduced to 6-dimensions. The coordinates of cell shape in this low-dimensional space, especially in the plane of the first and second shape modes, can discriminate migrator vs. non-migrator. We acquired an additional set of cell migration data with exactly the same experimental settings as a validation dataset. We

Text

showed that the prediction accuracy for NIH3T3 (5394 frames from 87 cells) is 97% for migrators and 95% for non-migrator; the prediction accuracy for LN229 (2790 frames from 45 cells) is 98% for migrators and 90% for non-migrators.

**Spatiotemporal persistence determines cell migration dynamics**

To explain why a single shape can describe migration dynamics, we explored how cell shape is related to the spatiotemporal persistence of migration dynamics. With the assumption of an equilibrium cell shape (10), the coordinates in the shape-space can indicate the potential persistence in shape deformation. As a cell changes its shape, the displacement of a membrane point is coupled to its neighboring points both spatially and temporally. To investigate how cells spatiotemporally coordinate the shape deformation in terms of contraction and protrusion, and the deformation dynamics leads to migration dynamics, we analyzed the spatiotemporal correlation using both short-term (30 minutes) and long-term (10 hours) cell membrane displacement data. The spatiotemporal correlation, $C_\mu(r\Delta\theta, s\Delta t)$, measures the lack of independence between two membrane points $\mu(\theta_i, t_i)$ and $\mu(\theta_i + r\Delta\theta, t_i + s\Delta t)$. A detailed description can be found in **Supporting Information**.

Two membrane points have a positive correlation when they both show positive (or negative) displacements, and vice versa. We first use 30-minute migration data to clearly identify migrator and non-migrator and their characteristics, then use 10-hour long migration data to address if the cell migration pattern can change over time.

Taking a typical migrator and a typical non-migrator (NIH3T3) and using only 30 minutes of their migration data, we showed that their spatiotemporal correlation functions are drastically different. The migrator has much larger negative and positive correlation spatiotemporally (**Figs. 3***A-C*), as compared with the non-migrator (**Figs. 3***D-F*). For the migrator (**Fig. 3***A*), the spatial correlation shows about half of the cell is positively correlated, while the other half is negatively correlated – this observation is more clearly indicated in the zeroth line contour plot (**Fig. 3***C*) – indicating that half of the cell is protruding while the other half is contracting, the most efficient form of migration with one single protrusion. The spatial correlation for the non-migrator, on the other hand, shows no large-scale correlation (**Fig. 3***D*) and numerous transitions between positive and negative correlation. Since the cell membrane deformation is bi-directional, inward and outward, positive correlation in time corresponds to persistent protrusion or contraction, and negative correlation in space means a large area of the membrane deform in the opposite direction, corresponding to coordinated protrusion and contraction. Negative correlation therefore is important for directional migration (11). The frequent transition between negative and positive correlation for non-migrators signifies the lack of both spatial and temporal persistence. On the contrary, two clearly distinct regions of positive correlation of the correlation map (**Fig. 3***B* and **Fig. 3***C*) and a negative correlation region characterize a migrator. The width of the negative correction in space is roughly $\pi$, which is the distance between the "front" and "back" of cell along the circular map.

**Discussion**

We next analyzed the spatiotemporal correlation of long-term (10 hours) migration. From the cell trajectories, it is clear that in addition to migrator and non-migrator, many cells transition between migration and resting, confirming that cell migration is a plastic, dynamical process. We chose cells with distinct migration behaviors to show spatiotemporal signatures of cell migration in terms of shape deformation. We showed that the correlation magnitude is not a feature for distinguishing migrator vs. non-migrator due to time averaging. The more persistent the migration, the less frequent transitions between positive and negative correlation (**Figs. 4***G-I*).

Text

The temporal persistence length corresponds to the duration of cell migration (**Fig. 4G**). For both the persistent migrator and the intermittent migrator, the width of negative correlation is smaller than π/2. We also observed that migrator has spatial correlation width greater than π/2. To further test the width threshold of the negative correlation region, we analyzed data of a cell that went through the process of polarization and initiation of directional migration. The correlation contour (**Figs. 4G-I**) shows that the width of negative correlation region increased from 0 to π. We also verified that the duration for the cell directional migrating equals to the temporal length of negative correlation region that is wider than π/2. Further detailed analysis of the spatiotemporal contour plots revealed more subtle migration dynamics, for example turning (**Fig. S5**).

**Discussion**

In summary, we analyzed quantitatively over a million images of cell morphology to determine if cell shape can discriminate cell migration behavior. The answer is a clear yes – when projected into the low-dimensional shape-space, the migrators and the non-migrators can be segregated. The discovery that a low-dimensional (6-dimensions) space accounts for 99% of all cell shape variation, for both mouse fibroblast and human glioblastoma cells, is a surprise. Even the simple stereotypical, highly persistent 'fan' shape of keratocytes requires 4 dimensions for 93% of the cell shape variation (4). Furthermore, the remarkable resemblance of the shape-spaces between the mouse fibroblast and human glioblastoma suggests the possibility of a combined common shape-space for the two cell lines. The implication is that, despite the vast genetic and phenotypic differences between the two cell lines, their morphologies during migration share enough similarities such that the differences are only reflected in the distributions in the space. This idea is tantalizing: can we use the differences in the shape-space to further distinguish different cells? Only a large number of morphology data with different cell

Text

lines can answer this question. Complex cell shapes and rich behaviors can emerge from interactions across different scales. Our study highlights that the spatial-temporal coordination of these interactions is essential for understanding the mechanism of cell behavior and shape determination.

**Methods**

Mouse embryo fibroblast NIH 3T3 cells and human glioma LN229 CDC42-GFP cells (gift from Dr. Erwin G. Van Meir, Emory University, Atlanta, GA) migrating on 2D fibronectin coated polyacrylamide gels with shear elastic modulus of 8.6 $kPa$ were imaged every 1 min for up to 10 hours using epifluorescence and confocal microscopy. Single cell migration data is obtained through manual cropping then thresholded using ImageJ (1.51n; National Institutes of Health). Analysis of cell outline, trajectories, MSD, directionality ratio and two-dimensional correlation were conducted using MATLAB (The MathWorks Inc., Natick, Massachusetts, 9.1 (R2016b) Linux (64-bit)). Experimental for cell culture, live cell imaging and computational procedures of cell membrane displacement, cell trajectories, cell major shape modes and joint probability density analysis, are given in **Supporting Information**.

Text

**Acknowledgments:**

We thank Integrated Cellular Imaging (ICI) Shared Resource of Winship Cancer Institute of Emory University. We are also grateful to Frank Hu for assistance with experiments and data analysis. XH was partially supported by Molecular Basis of Disease Fellowship at Georgia State University. YJ acknowledges funding support from NIH R01CA201340 & R01EY028450. KC and NF acknowledge funding support from NIH R01GM115763. Authors declare no competing interests.

**Author contributions:**

XH &YJ designed the analysis. XH, YJ and KC designed the experiments. XH & KC performed cell experiments, KC and XH did the epifluorescence imaging, XH did the confocal imaging and the data analysis. XH wrote the first manuscript with YJ. XH and KC wrote the Supplementary Information with YJ. XH, YJ, KC and NF edited manuscript.

**Materials & Correspondence:**

Correspondence and material requests should be addressed to YJ.

**Competing interests:**

No.

Text

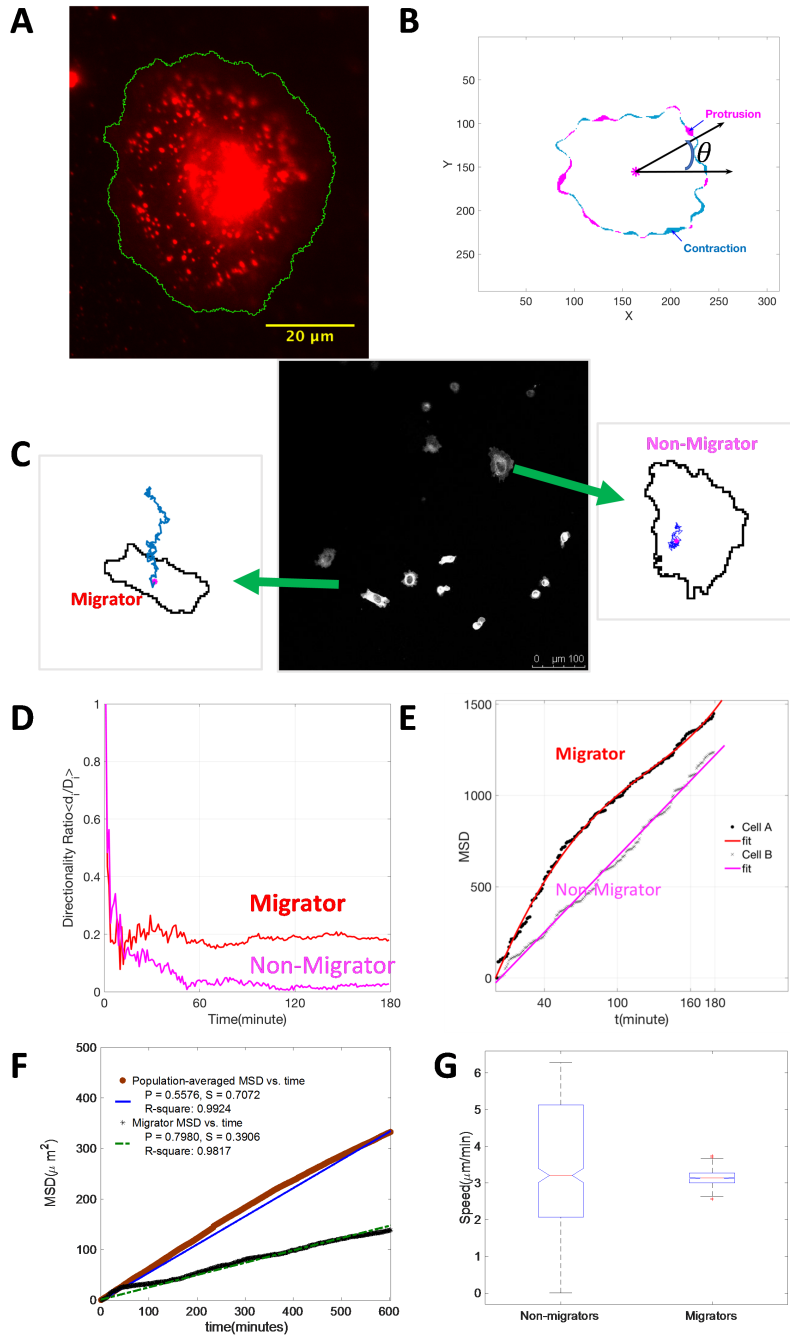

**Fig. 1. Cellular migration persistence and speed variation on a 2D surface.** (**A**) Fluorescence image of a LN229 cell expressing CDC42-GFP (shown in red) on a fibronectin-coated polyacrylamide gel with the cell outline (green). (**B**) Cell shape deformation $\mu(\theta, t)$ is the difference of cell shape between two consecutive time frames $t + \Delta t$ and $t$, measured as a radial displacement from centroid to the cell outline: protrusion is positive (magenta) and contraction is negative (blue). (**C**) From a frame of a LN229 cells, we illustrate a migrator and a non-migrator cell as outlines (at $t = 0$) with trajectories (over $3\ hrs$); (**D**) The directionality ratio ($DR$) and (**E**) Mean square displacement (MSD) over 3 hours of the migrator (red) and the non-migrator (magenta). (**F-G**) Population average of MSD and cell speed variation for migrators ($n = 130$) and non-migrators ($n = 897$).

Text

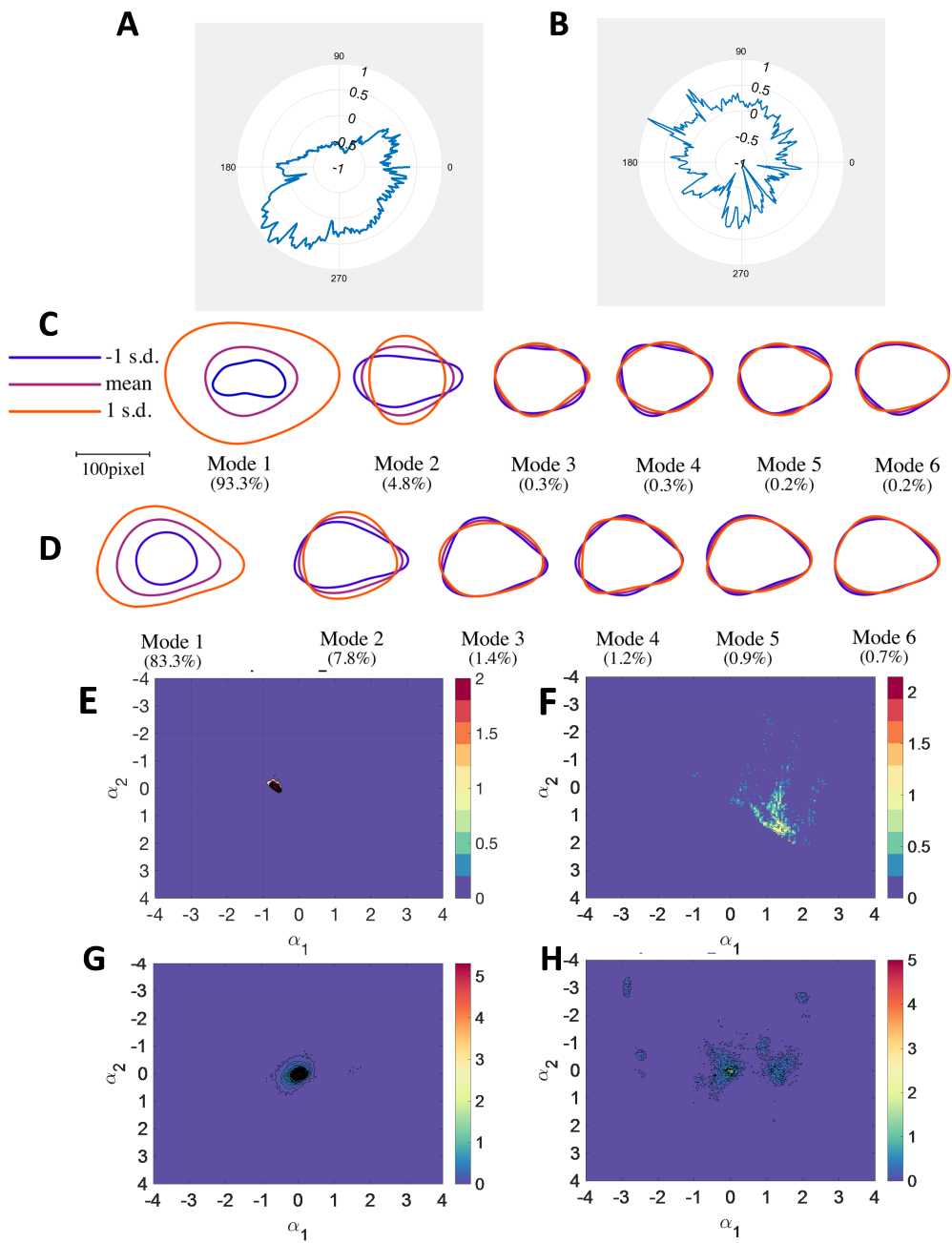

**Fig. 2. Cell shape variation, shape-space and coordinates.** (***A-B***) The membrane displacement $\mu(\theta, t)$ for a migrator (***A***) shows larger amplitudes and a more biased angular distribution than that for a non-migrator. (***B***) The concentric circles mark $-100\%, -50\%, 0\%, 50\%,$ and $100\%$ displacement. (***C-D***) Cell shape variation modes $\eta_i$ of NIH3T3 (***C***) and LN229 (***D***) and the respective percentages of cell shape variation captured by each mode. Scale bar is $200\ \mu m$. The first 6 modes account for 99% of shape variance, forming a 6-dimensional cell shape-space, for both NIH3T3 and LN229. The coordinates of a cell shape in this shape-space projected in the first two shape-modes are $(\alpha_1, \alpha_2)$. (***E-H***) The joint probability density $\rho(\alpha_1, \alpha_2)$ for all non-migrators is confined in a small region near the origin for both NIH3T3 (***E***) and for LN229 (***G***), while the joint probability density for all migrators resides in much wider regions for NIH3T3 (***F***) and LN229 (***H***). Each data point corresponds to one cell image.

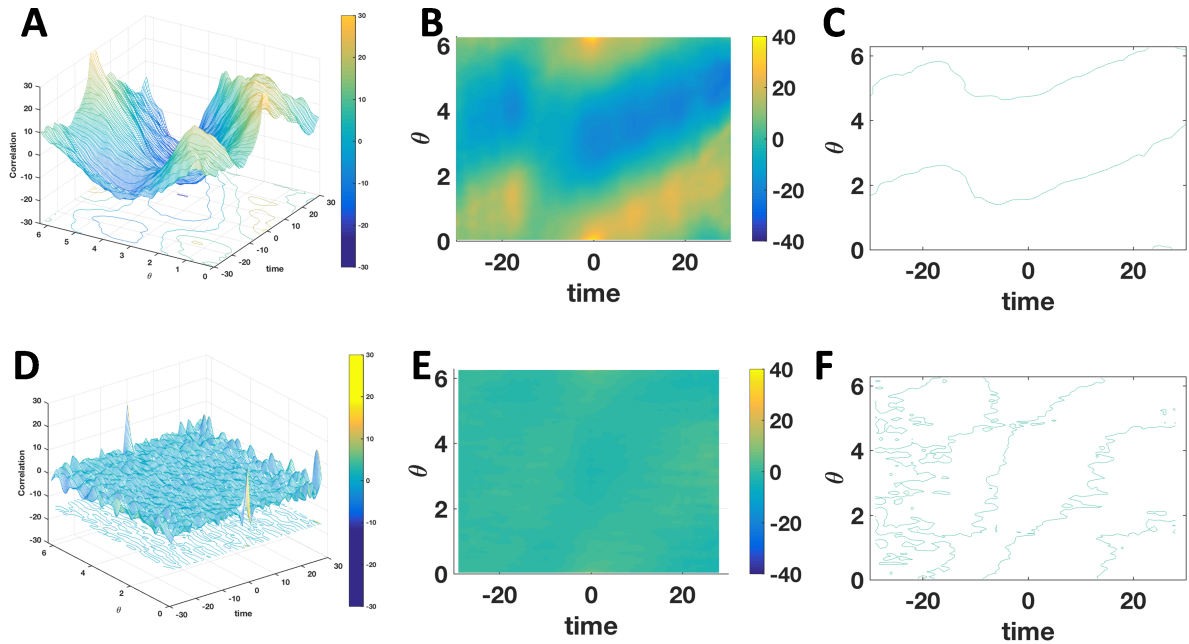

**Fig. 3. Spatiotemporal correlation analysis of short-term (30 minutes) cell membrane displacement for a migrating and a non-migrating NIH3T3 cells.** (*A-C*) The spatiotemporal correlation of a migrator as a surface plot, a heat-map, and a contour plot. (*D-F*) The spatiotemporal correlation of a non-migrator as a surface plot, a heat-map, and a contour plot. The zeroth value contour lines (*C* and *F*) mark the transition between positive and negative correlation.

Text

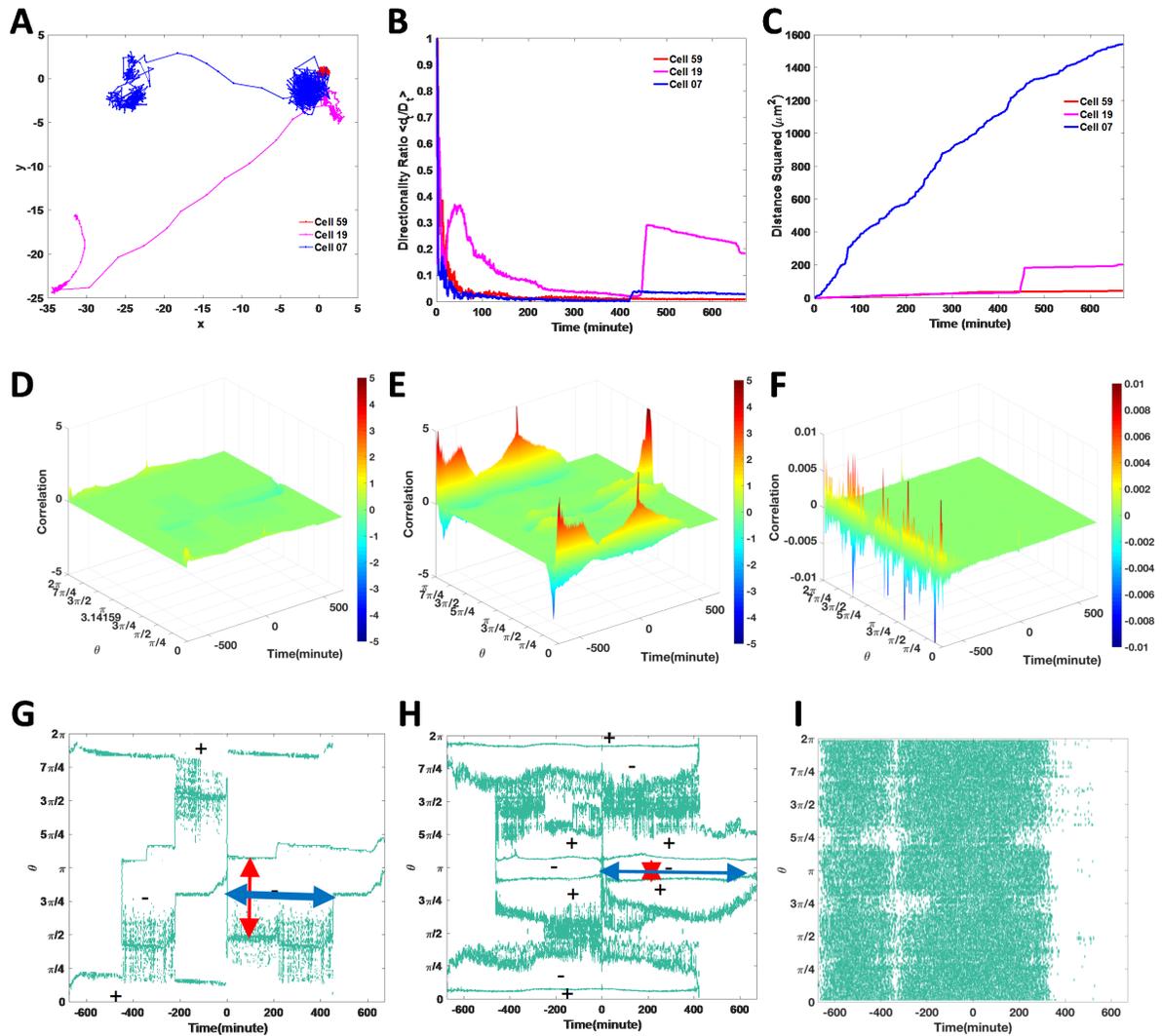

**Fig. 4. Distinct features of migration behavior.** Cell trajectories (**A**), directionality ratio (**B**) and mean square displacement (**C**) of migrator (magenta, Cell 19), non-migrator (red, Cell 59) and intermittent (blue, Cell 07) on 2D substrates in absence of asymmetric signal. Cells 59, 19 and 07 are chosen from NIH3T3 to represent three types of migration behavior (**Fig. S2**). Spatiotemporal correlation and contour plots of migrator (**D** and **G**), intermittent (**E** and **H**) and non-migrator (**F** and **I**). Contour plots (**G, H** and **I**) were obtained from (**D-F**) to show the transition between positive (+) and negative (−) correlation. Red arrow indicates spatial scale of negative persistence, which is greater than $\pi/2$ for migrator (**G**) and less than $\pi/2$ for intermittent (**H**). Blue arrows indicate the length of negative correlation in time.

# Supplementary Information for

"Coordinates in Low-Dimensional Cell Shape-space Discriminate Migration Dynamics from Single Static Cell Images"

Xiuxiu He[1], Kuangcai Chen[2], Ning Fang[2], Yi Jiang[1*]

[1] Department of Mathematics and Statistics, Georgia State University, 14th Floor, 25 Park Place, GA, Atlanta, USA 30303-3083.

[2] Department of Chemistry, Georgia State University, 3rd Floor, 145 Piedmont Ave SE, Atlanta, GA, USA 30303.
[*] Corresponding author. Email: yjiang12@gsu.edu.

We provide detailed descriptions on experimental methods and data analysis.

Supplementary Materials and Methods

**Supplementary Materials and Methods**

**Substrate Preparation**

Published protocols were followed with certain modifications to prepare the extracellular matrix (ECM) coated polyacrylamide (PAA) gel on coverslips for cell culture and imaging of cell migration(1). Briefly, cleaned 22 × 22 mm glass coverslips were first activated with 2% (v/v) 3-aminopropyltrimethoxysilane (281778-100ML, Sigma Aldrich). Working solution containing the final concentrations of 7.5% acrylamide (161-0140, Bio-Rad) and 0.3% bis-acrylamide (BP1404, Fisher Scientific) in Milli-Q water (296.75 µL) for making 8.6 kPa PAA gels were added between Rain-X wiped hydrophobic glass slide surface and amine activated coverslips. Coverslips with PAA gels attached were removed from the microscope slide surface upon the completion of the PAA polymerization (~10 min at room temperature) initiated by TEMED (BP150-20, Fisher Scientific) and 10% (w/v) Ammonium Persulfate (BP179-100, Fisher Scientific) and kept hydrated in water. The surface of the PAA gel on the coverslips were activated by incubation with hydrazine hydrate (225819-250G, Sigma Aldrich) to cross-link with sodium meta-periodate (20504, Thermo Fisher) oxidized Fibronectin (FN, 33016015, Thermo Fisher). These FN coated PAA gel attached coverslips were then maintained hydrated in 1× phosphate buffer saline (pH 7.4, 10010023, Thermo Fisher) before being used for cell plating.

**Cell Culture**

Complete cell culture medium was made of high glucose Dulbecco's Modified Eagle's Media (DMEM, 10-013-CV, Corning) with 10% fetal bovine serum (FBS, 26-140-079, Fisher Scientific) and 1% penicillin/streptomycin solution (30-001-CI, Corning). Both NIH3T3 mouse fibroblast cell line and doxycycline inducible GFP-LN229 cell line (gift from Dr. Van Meir, Emory University,

Atlanta, GA) were cultured with complete cell culture medium in T25 flasks under standard humidified culture condition in a 37 °C incubator with 5% $CO_2$ prior to plating the cells onto FN coated PAA gel attached coverslips. A final concentration of 5 µM DiI (D282, Thermo Fisher) was used to label NIH3T3 cells according to manufacturer's instructions. Cells were plated and incubated for 6 h and then washed with fresh medium before imaging. Cells were observed to divide normally during and after the 10 h imaging.

**Microscopy and Live Cell Imaging**

Time-lapse epifluorescence imaging of the cell migrations of DiI labeled NIH3T3 cells and doxycycline induced GFP-LN229 cells on FN coated PAA gel surface was carried out on a Nikon Eclipse 80i upright microscope equipped with a mercury lamp, heating stage, Nikon (Melville, NY) Plan Fluor 20× 0.5 NA objective, Hamamatsu ORCA-Flash 4.0 V2 COMS camera. The heating stage was set to 37 °C. A homemade program was used to control a shutter of the mercury lamp to reduce photobleaching and Micro-Manager was used to control hardware and to take time-lapse image with 1 min interval. Live cell fluorescence images were also obtained on a Leica (Wetzlar, Germany) SP8 inverted microscope with a confocal galvonometric scanner, motorized stage, mercury lamp, heating state, Argon laser with 458 nm, 488 nm, and 514 nm wavelengths. Leica application Suite-Advanced Fluorescent software was used to control acquisition. Images were acquired every minute for up to 10 h using HC PL APO 20x/0.75 CS2 air WD 0.62 mm objective.

**Cell Shape Extraction**

Single cell image stacks were extracted from the NIH3T3 and LN229 live cell imaging data sets by manual cropping the non-overlapping cells for the migration durations. Homemade macros with ImageJ (1.51n; National Institutes of Health) were used to threshold every single cell image, creating static frames of cell boundaries as text files. Cell shape sequences were then computed using Matlab (R2016b) Linux (64-bit). We have analyzed 517 NIH3T3 cells and 510 LN229 cells, 373,796 frames in total; for validation, we used additional 87 NIH3T3 cells (5394 frames) and 45 LN229 cells (2790 frames).

**Analysis of Cell Shape and Motion**

Montage image of each cell was processed using Matlab (The MathWorks Inc., Natick, Massachusetts, 9.1 (R2016b) Linux (64-bit)) to obtain cell shape dynamics data. Shape data $\Omega(X_{i\Delta t}, Y_{\Delta t})$ were obtained using the text image of each cell in the Cartesian system set by the first frame, where $(X_{i\Delta t}, Y_{\Delta t})$ represents the coordinates of cell contour, and the centroid $c(i\Delta t) = (x_{i\Delta t}, y_{i\Delta t})$ was calculated for each time point. Further we defined the instantaneous cell speed as the displacement of the centroid, $\vec{v}(t) = \frac{c(t+\Delta t) - c(t)}{\Delta t}$ at time interval of $\Delta t$ equals to 1 minute.

The mean square displacement (MSD) of a cell is defined as:

$$MSD(n\Delta t) = \frac{1}{n}\sum_{i=1}^{n}\left[\left(x_{i\Delta t} - x_{(i-1)\Delta t}\right)^2 + \left(y_{i\Delta t} - y_{(i-1)\Delta t}\right)^2\right], \text{(S.1)}$$

where $n$ is the total number of frames for the cell of interest. We fit the MSD data with a persistent random walk model:

$$MSD(\Delta t) = 2P^2 S^2 \left[\frac{\Delta t}{P} - 1 + e^{-\frac{\Delta t}{P}}\right] + 4\sigma^2, \text{(S.2)}$$

where $P$ is the persistent time, and $S$ is the instantaneous cell speed. We understand the differences in $P$ and $S$ for migrators and non-migrators using the stochastic differential equation for cell velocity, v, without external bias(2):

$$dv(t) = -\beta v(t) + \sqrt{\alpha}dW(t), \text{(S.3)}$$

where $\beta$ is the decay rate of the velocity, W(t) is the Weiner process with a magnitude $\alpha$. The persistent time and the cell speed can be derived as $S = \sqrt{\alpha/\beta}$ and $P = 1/\beta$. We showed that migrators have longer persistent time and the non-migrators have greater magnitude of the random fluctuation in speed (**Fig. 1I**).

To analyze the cell shape dynamics at the subcellular level, we used the circular mapping, the centroid was defined as the origin $O$, the radial distance from $O$ to the cell contour were calculated for each frame of image. We further represented the cell shape as $\Omega(\theta, t)$, the radial distance from $O$ to the cell contour at the direction $\theta$, with $\theta = 0°$ being fixed in the right-hand direction on the horizontal axis as in the Cartesian system. Based on which we computed cell shape deformation $\mu(\theta, t)$ as change of the radial distance in direction $\theta$. We finely discretized the cell contour using $\theta$ for spatial resolution and there was no rotational cell membrane segment motion being considered. This setting is sufficient for the live cell imaging data that we obtained, noticed that other approached might be necessary for more complicated cell shape deformation.

**Shape Modes Analysis**

Cell contour was extracted as intensity isoline from each static frame of cell image using Celltool (3). The contours were then aligned along their long axes to eliminate pose and relative position. To decompose this space into a basis set of orthogonal "shape modes", principal components analysis was then performed on a large population of single cell images, for each cell line respectively. The obtained major modes $\vec{\phi_i}(\theta)$ were ranked by the percentage of variation that they captured.

**Prediction Accuracy**

The images are acquired for 61 minutes at the rate of one minute per frame. We collected 45 LN229 cells (2790 frames) and 10 (620 frames) of which are migrators. For NIH3T3, we collected 87 cells (5394 frames) in total, and 22 cells (1364 frames) are migrators. Each frame is treated as an identical data point. We first extracted cell contours (10939 for NIH3T3, 5760 for LN229) from outlies that we obtained using ImageJ (1.51n; National Institutes of Health). And then $\alpha_1$ and $\alpha_2$ were calculated for each contour using the cell shape variation modes (**Fig. 2C** and **Fig. 2D**). The prediction accuracy is calculated, for migrator and non-migrators separately, as the number of cell frame correctly identified divided by the total number of single cell frames.

**Joint Probability Density**

Cell shape coordinates in the shape-space from each cell static frame of cell image was approximated using major cell shape variation modes. The magnitudes $\alpha_i(t)$ were used as data for discrete random variables $\alpha_i$ assuming that

$$f_{\alpha_i,\alpha_j}\left(\alpha_i(t),\alpha_j(t)\right) = f_{\alpha_i|\alpha_j} P\left(\alpha_j = \alpha_j(t)\right) = P\left(\alpha_i = \alpha_i(t)| \alpha_j = \alpha_j(t)\right) f_{\alpha_i}(\alpha_i(t)). \text{(S.4)}$$

Joint probability density $\rho(\alpha_i, \alpha_j)$ was computed for all combinations of $(i, j)$ with $i \neq j$.

**Spatiotemporal Correlation**

Cell shape deformation $\mu(\theta, t)$ was discretized in space and time. $\Delta\theta = 2\pi/N$, $\Delta t$ is one minute and total time $T = (M + 1)\Delta t$. The spatiotemporal correlation was computed as:

$$C_\mu(s\Delta\theta, r\Delta t) = \begin{cases} \frac{1}{N(M-s)} \sum_n \sum_{m=1}^{M-s} \mu_{n,m} \cdot \mu_{n+r,m+s}, s \geq 0 \\ \frac{1}{N(M+s)} \sum_n \sum_{m=1}^{M+s} \mu_{n,m} \cdot \mu_{n+r,m+s}, s < 0 \end{cases} \quad (S.5)$$

We chose not to remove the mean from cell shape deformation $\mu(\theta, t)$ for the following reasons. First, mean value is not biologically meaningful. $E[\mu(\theta, t)]$ over $\theta$ is the instantaneous speed at $t$, which varies in time. Second, $\mu(\theta, t)$ is also displacement of membrane segment in the angle $\theta$, so mean displacement is not useful. Removing the mean $E[\mu(\theta, t)]_{(\theta,t)}$ only scales down correlation magnitude.

**Equation of motion and deformation**

Cell motility is traditionally characterized by cellular level parameters, such as velocity and directional persistence of the centroid of the cell. The centroid of a cell performing a persistent random walk can be described as a Langevin equation. We assume that each bit of the membrane, called a membrane element, also follows a generalized Langevin equation (2):

$$dv_i(t) = -\vec{\beta}_i^T \vec{v}(t) dt + \alpha_i^T dW_i(t), \quad (S.6)$$

where the velocity of each cell membrane element $v_i$ is affected by the deterministic resistance to the element's motion and the random fluctuations $W_i(t)$ with amplitude $\alpha_i$. Given a time series of a cell, we can fit the motion of every element in this form. The matrix, $B = \{\beta_{i,j}\}$, characterizes the temporal relations of the displacements between membrane elements: $B(i, i) < 0$ corresponds to contraction of the $i^{\text{th}}$ membrane element, while $B(i, i) > 0$ corresponds to an extension; $B(i, j) > 0$ implies that the $i$ th and $j$ th membrane elements move in the same direction, and vice versa. This matrix is the effective interaction network amongst all membrane elements, the interactions include contributions from the cytoskeleton, membrane tension, myosin contraction, and all other possible interactions including fluid pressure or flow.

Each cell shape has the coordinate $(x_1, x_2, \ldots x_K)$ in the $K$-dimensional orthogonal shape space $\Phi: \Omega(\theta_i) = \sum_{j=1}^{K} x_j(t)\varphi_j(\theta_i)$. Replacing $v_i(t)$ with $d\Omega(\theta_i)/dt$, we have

$$-\vec{\beta_i}^T \sum_{j=1}^{K} dx_j(t)\varphi_j + \alpha_i^T dW_i(t) = \sum_{j=1}^{K} d^2 x_j(t)\varphi_j(\theta_i). \quad \text{(S.7)}$$

This equation of motion describes the shape and migration dynamics in terms of the effective interaction network amongst membrane elements. It connects cell shape dynamics and cell movement behavior.

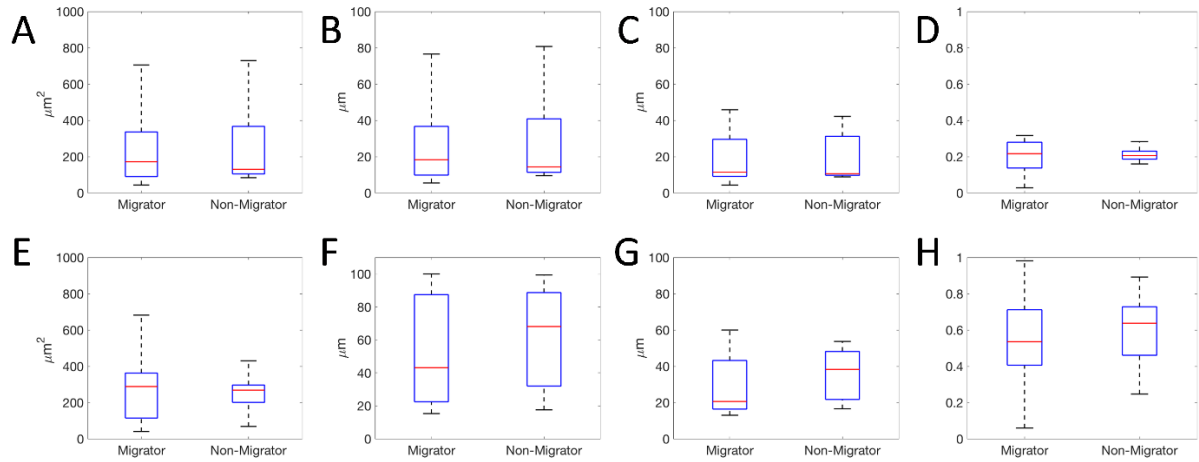

**Supplemental Figure 1. Simple shape measurements do not show significant differences between migrator and non-migrator.** Top (**A-D**): LN229; Bottom (**E-H**): NIH3T3. (**A** and **E**) Cell area. (**B** and **F**) Major axis. (**C** and **G**) Minor axis. (**D** and **H**) Eccentricity.

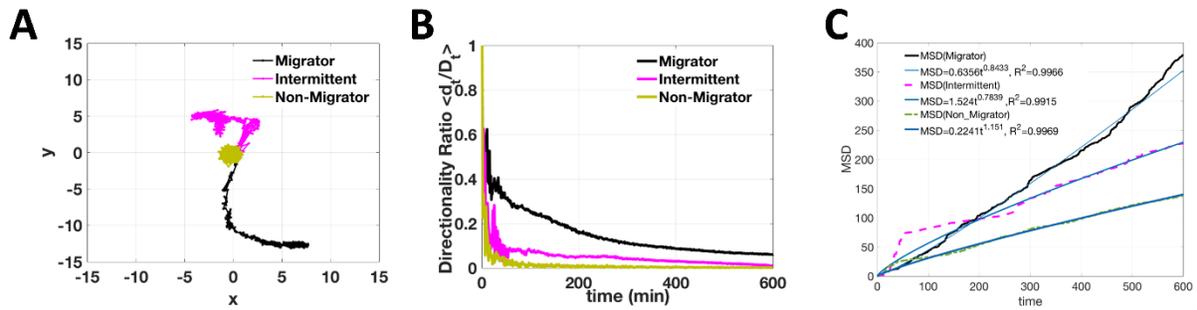

**Supplemental Figure 2. Distinct features of migration behavior.** (**A**) Cell trajectories (LN229) show distinct migration behaviors. (**B**) Directionality ratio discriminates migrator, non-migrator and intermittent. (**C**) MSD was fit to $MSD(t) = Dt^\alpha$, where $D$ is diffusion coefficient. The parameters $D$ and $\alpha$ did not show a consistent trend to discriminate migration behaviors.

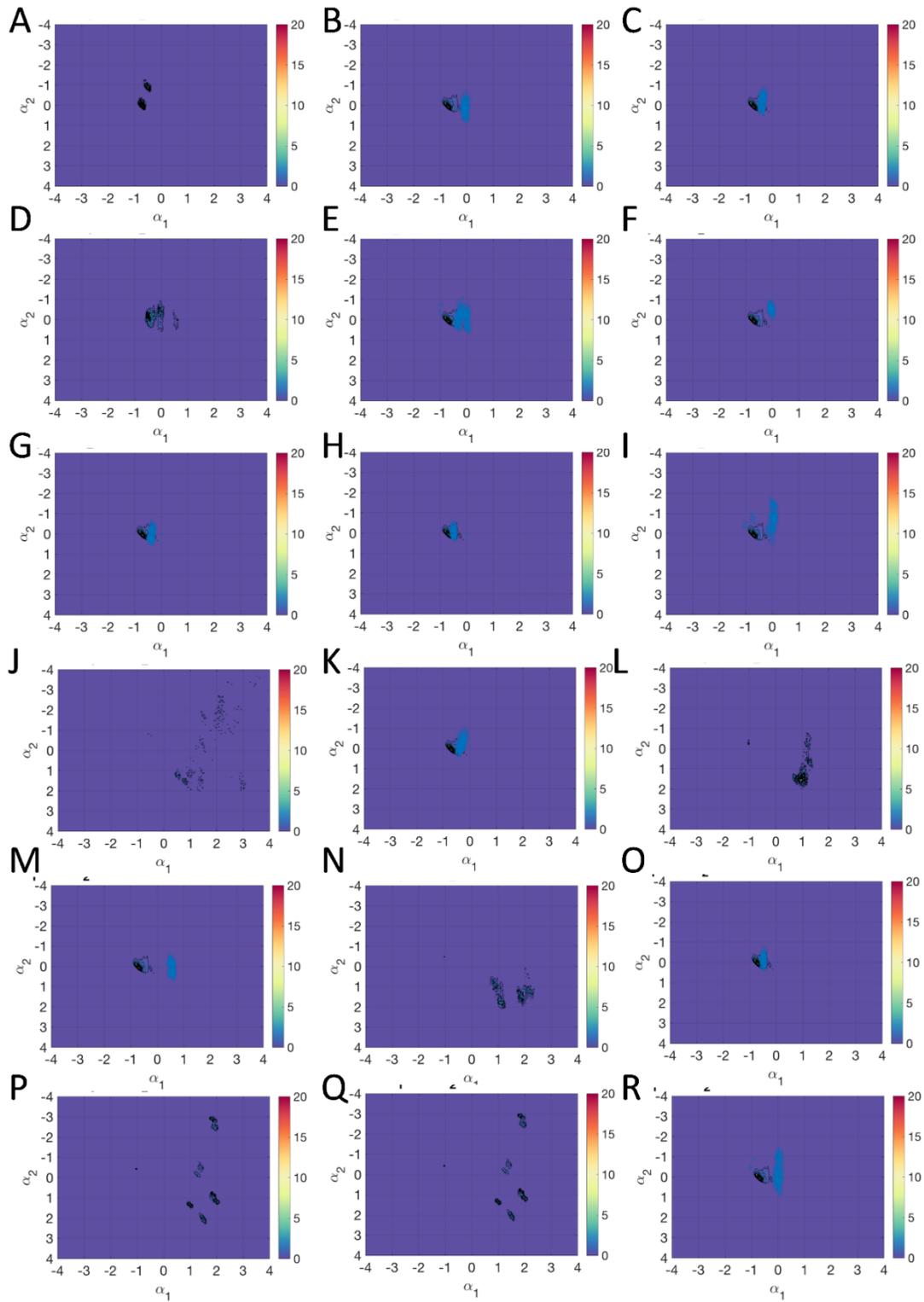

**Supplemental Figure 3. Coordinates ($\alpha_1, \alpha_2$) of migrators (LN229) that overlap with the non-migrator region. (A, D, J, L, N, P,** and **Q)** only show migrator cells. (**B, C, E, F, G, H, I, K, M, O,** and **R)** Migrators are highlighted using blue + markers.

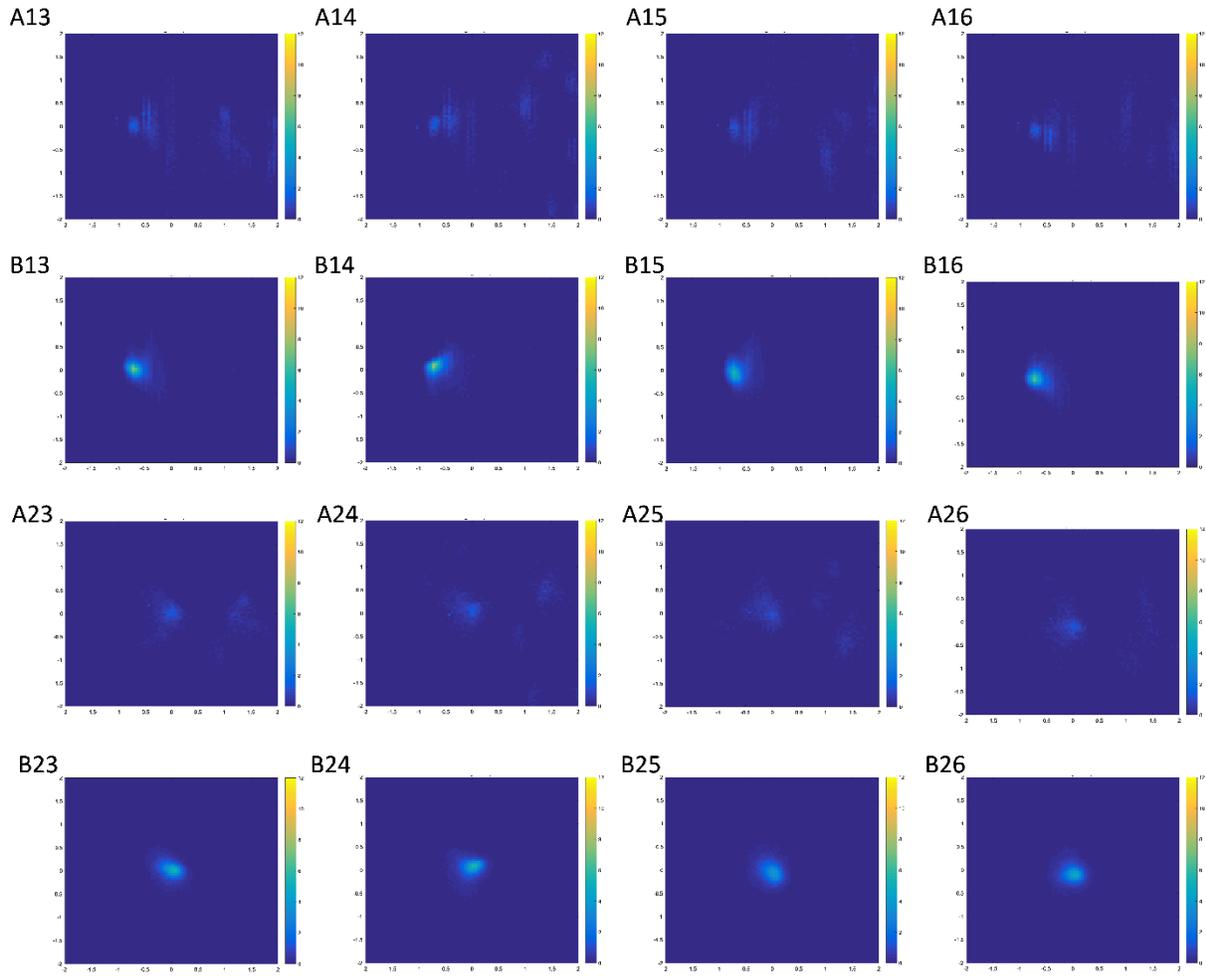

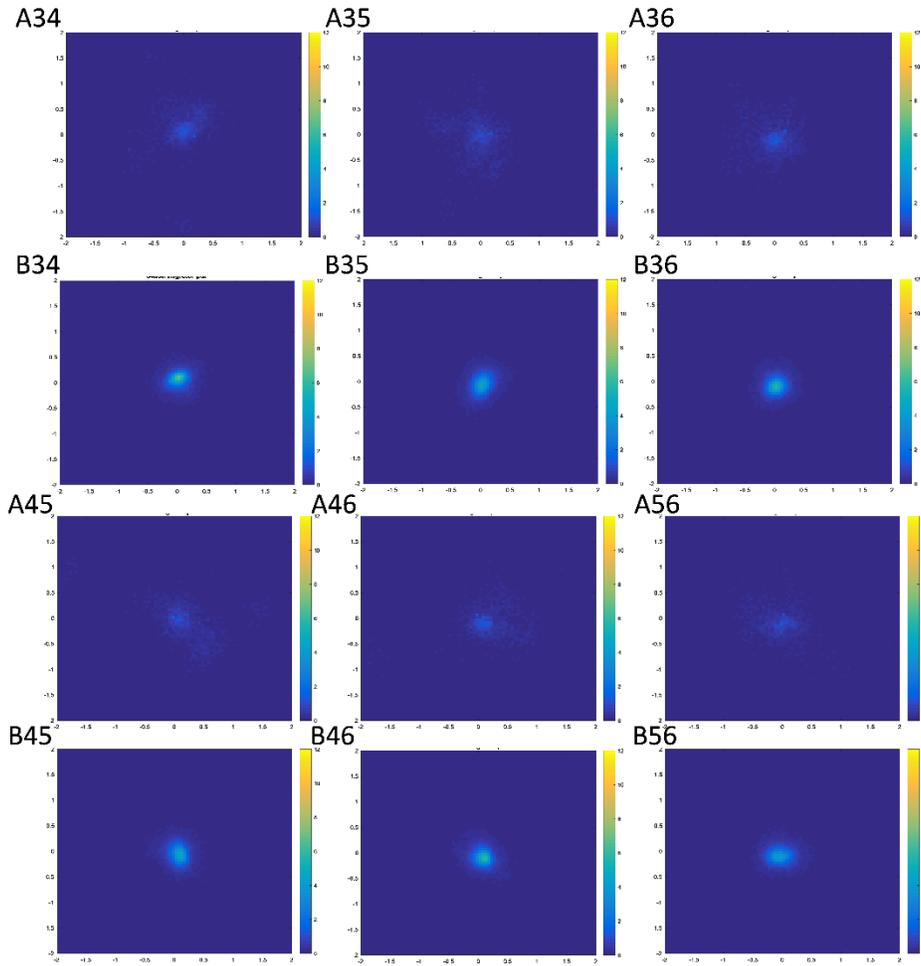

**Supplemental Figure 4. Combinations of amplitudes corresponding to major shape modes that capture lower percentage of variation, {*i, j*}.** $a_{i,j}$, The joint probability density $\rho(\alpha_i, \alpha_j)$ for coordinate pairs i and j for migrators. $b_{i,j}$, The joint probability density for non-migrators.

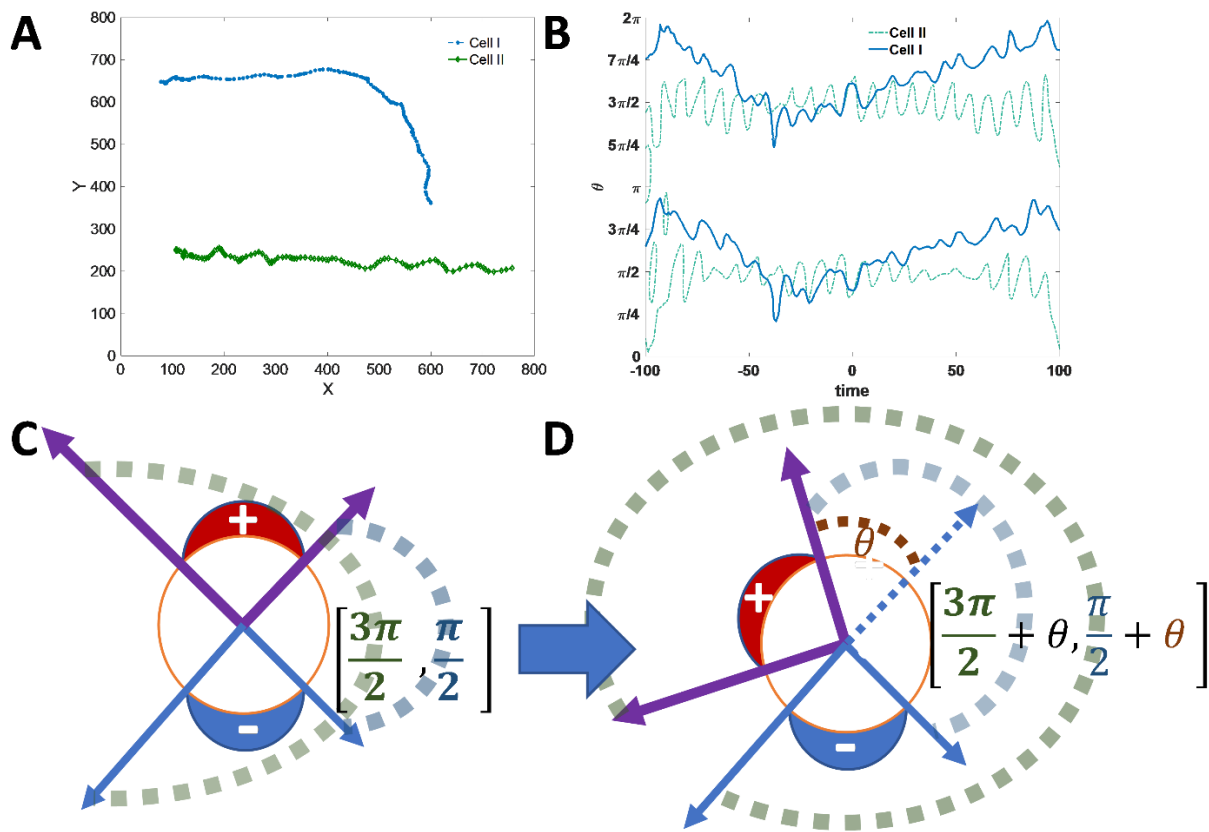

**Supplemental Figure 5. Turning during migration**. (**A**) Cell trajectories for two migrators. (**B**) Contour plot of spatial-temporal correlation. (**C**) Cell with a persistent front protrusion (red with +) and rear contraction (blue with −), which resulted in negative correlation between $\left[\frac{3\pi}{2}, \frac{\pi}{2}\right]$, consistent with **B**. (**D**) As the cell turns, the distance between protrusion and contraction increase by angle $\theta$ (brown), which is reflected as the negative correlation region shifted to $\left[\frac{\pi}{2}+\theta, \frac{3\pi}{2}+\theta\right]$.

**Supplementary References**